\documentclass[reprint,secnumarabic,amssymb,amsmath,nobibnotes,aps,prd] {revtex4-1}
\usepackage[all, knot]{xy}

\newtheorem{theorem}{Theorem}

\newtheorem{lemma}{Lemma}
\newtheorem{corollary}{Corollary}

\usepackage[all, knot]{}
\begin{document}

\title{Towards superconformal and quasi-modular representation of exotic smooth $\mathbb{R}^4$ from superstring theory II}

\author{Torsten Asselmeyer-Maluga}

\email[REV\TeX{} Support: ]{torsten.asselmeyer-maluga@dlr.de}

\affiliation{German Aero space center, Rutherfordstr. 2, 12489 Berlin}

\author{Jerzy Kr\'ol}

\email[REV\TeX{} Support: ]{ iriking@wp.pl}

\affiliation{University of Silesia, Institute of Physics, ul. Uniwesytecka 4,
40-007 Katowice}

\begin{abstract}
This is the second part of the work where quasi-modular forms emerge from small exotic smooth $\mathbb{R}^4$'s grouped in a fixed radial family. $SU(2)$ Seiberg-Witten theory when formulated on exotic $\mathbb{R}^4$ from the radial family, in special foliated topological limit can be described as $SU(2)$ Seiberg-Witten theory on flat standard $\mathbb{R}^4$ with the gravitational corrections derived from coupling to ${\cal N}=2$ supergravity. 

Formally, quasi-modular expressions which follow the Connes-Moscovici construction of the universal Godbillon-Vey class of the codimension-1 foliation, are related to topological correlation functions of superstring theory compactified on special Callabi-Yau manifolds. These string correlation functions, in turn, generate Seiberg-Witten prepotential and the couplings of Seiberg-Witten theory to ${\cal N}=2$ supergravity sector.  Exotic 4-spaces are conjectured to serve as a link between supersymmetric and non-supersymmetric Yang-Mills theories in dimension 4. 
\end{abstract}


\maketitle

\section{Introduction}
Since superstring theory provides very rich mathematics which, by now, is not yet fully appreciated, it was proposed to use the \emph{supersymmetric and conformal} tools of superstrings as suitable for grasping several aspects of exotic smoothness on open 4-manifolds.\cite{AssKrol2010ICM,AsselmKrol2011d,AsselmeyerKrol2011,AsselmeyerKrol2011b,QG-2012,Krol2011d,Krol2010} 
In the first part of this work we made use of the correlation functions of superconformal algebras with ${\cal N}=4$ supersymmetries as expressing invariants of some topological ingredients of the structure of small exotic smooth $\mathbb{R}^4$. The whole approach is based on the standard by now calculating the topological invariants, like Reshetikhin-Turaev-Witten or Chern-Simons,  of 3-spaces from ${\cal N}=4$ and ${\cal N}=2$ superconformal algebras. The 3-manifolds are homology 3-spheres and appear as valid ingredients of the construction of exotic 4-structures. This gives only partial understanding of open 4-smoothness, though indicates on topological origins of some algebraic constructions, like algebraic ends of exotic $\mathbb{R}^4$. Such, quite formal approach emphasizes however, the special role played by quasi-modular forms and Mock theta functions in approaching exotic 4-spaces, though the role seemed to be justified only partially and rather accidentally.

In the present part we approach small exotic smooth $\mathbb{R}^4$'s from fixed radial family as 4-dimensional manifold which can be placed in string theory background, on the one hand, and can be an arena for formulating 4-d quantum field theory, on the other. In fact, we are interested in a version of supersymmetric Seiberg-Witten (SW) theory on exotic $\mathbb{R}^4$, $e$. This gives direct relation to topological type II superstring theory. Surprisingly, the quasi-modular expressions are crucial here and they appear as gravitational corrections in 6-d Kodaira-Spencer theory of gravity, as topological string amplitudes, and as the gravitational corrections to SW theory. However, since we do not know any metric on any exotic $e$ we are searching for a limit of Yang-Mills theory on $e$ where certain invariant data dominate. Exotic $\mathbb{R}^4$ are non-flat Euclidean smooth 4-manifolds which can be considered also as gravitational backgrounds. Thus, SW theory on flat $\mathbb{R}^4$ acquires gravitational corrections due to exotic $\mathbb{R}^4$. In the special limit considered in this paper, these corrections match supergravity ${\cal N}=2$ corrections to flat SW theory. The limit is not purely topological since all exotic $\mathbb{R}^4$ are topologically trivial. It is rather special \emph{foliated topological} limit where invariant data of some codimension-one foliations dominate.

To this end we are going to show that the universal Godbillon-Vey (GV) class of the codimension-one foliations of certain compact 3-manifold 
\begin{itemize}
\item[-] is assigned to exotic $e$ from fixed radial family (Sec. \ref{GV});
\item[-] modifies structure of the crossed product of the ring of modular forms ${\cal M}$ of arbitrary level by the group ${\rm GL}^+(2,\mathbb{Q})$, ${\cal M}\ltimes {\rm GL}^+(2,\mathbb{Q})$ (Sec. \ref{GV}); 
\item[-] gives rise to the gravitational corrections to SW theory.
\end{itemize}
These corrections to the effective SW Lagrangian are related to the correlation functions of the topological string theory compactified on local Callabi-Yau (CY) manifolds.
The connection of SW theory and string theory is very rich field and is based, among others, on mirror symmetry, special CY geometry and (quasi) modular forms again, and has been intensively studied (see, e.g. Refs. \cite{BCOV1993,Klemm1997,Nekr2002,Klemm2009}). 
 The key aspect of the correspondence is the quasi-modularity generated by Seiberg-Witten curve on the one hand, and monodromy of periods in CY, on the other. This is explained in the next section and some background material on SW theory is placed in the appendix. 

Note that the action of the radial family \emph{as a whole} over the modular ring of functions does not depend on the specific representant of the radial family. Hence, the radial family is, presumably, the entity of special importance in some limit of string theory. This particular situation when the result does not depend on specific metric of exotic $\mathbb{R}^4$ in the radial family, but depends on universal GV class of the foliations, indicating that the structure of the members is exotic, is what we called foliated topological limit above, and it is analyzed in Sec. \ref{Grav-Corr}.

We close the paper  with a bunch of conjectures relating the nature of a theory of quantum gravity (QG). There emerges the possibility that non-supersymmetric Yang-Mills theory, though on exotic $\mathbb{R}^4$, could be continuously related with SW theory on flat $\mathbb{R}^4$ where mass gap and confinement appear as the result of supersymmetry. More thorough analysis of these ideas will be presented in our separate publication.

\section{Quasi-modular forms in topological strings and 4-d Seiberg-Witten theory}\label{strings-1}
In this section we want to recall some facts explaining how the polynomials of second Eisenstein series emerge in topological string theory and in SW theory without, and with, massless flavor multiplets \cite{Klemm2009}. Our aim here is to understand how the quasi-modularity emerges and to approach its indeed universal power in the above theories. The presentation follows mainly the results of Refs. \cite{Klemm2006,Klemm2007b,Klemm2009,BCOV1993}.

SW theory is the low energy effective action for ${\cal N}=2$ 4-d gauge theory which, due to the holomorphicity, admits exact solutions. The Seiberg-Witten geometry of the moduli space determines Seiberg-Witten curve and the 1-form (SW differential) from which the solutions are derived (see also \ref{app}). 
Now, gauge theories with ${\cal N}=2$ supersymmetries can be embedded into type II superstring theory when it is compactified on non-compact Calabi-Yau manifolds (see, e.g. Ref. \cite{Klemm2007b}). The extracting 4-d field theory data from superstrings can be performed in various ways. Usually, the approach relies on the topological limit of type II superstrings on open, local CY manifolds and taking so called double scale limit, where gravity decouple, $M_{Pl}\to 0$, and stringy effects disappear, i.e. $g_s\to 0$. Then higher genus functions of the topological strings ${\cal F}_g(\underline{t})$, as the function of K{\"a}lher moduli $\underline{t}$, are considered in the above double scale limit. It appears that the limiting expressions are now the functions of SW moduli $a,a_D$ and are deeply related to superpotential and thus defines the SW theory. The core for such correspondence between 4-d ${\cal N}=2$ $SU(2)$ SW theory and topological strings on noncompact CY, is the structure of Riemann surface $C$ with the meromorphic differential $\lambda_{SW}$ with non-vanishing residua (\ref{app}). 
Such structure appears in both cases, in SW theory as the celebrated SW curve based on monodromies of the moduli space, and in the compactified topological string theory as based on monodromy of singularities of CY connected with the periods of the holomorphic 3-form $\Omega$. Then, the mathematical task, i.e. Riemann-Hilbert problem, to find multivalued functions $a$, $a_D$ on the moduli space of the curve with the prescribed monodromies and with ${\rm Im}\, \tau >0$, has the unique solution and the prepotential ${\cal F}$ can be determined from the K{\"a}lher geometry of the moduli. 

There are 2-types of topological string theories which, geometrically, rely on mirror symmetry. A-model depends on symplectic geometry of CY manifolds, while the B-model as defined on the mirror image of the A-model, depends on the deformations of the complex structure of the mirror CY. The remarkable feature of the topological string theory is the possibility to compute the non-perturbative $F$-terms in the effective supersymmetric gauge theories, like SW theory, emerging via compactifications. The $F$-terms of gravity theories can be computed this way. The gravitational corrections were derived from the topological string theory \cite{BCOV1993} which was defined as twisted $\sigma$-model on local CY, and coupled to a gravity theory on the target. Such a quantum theory of gravity is called Kodaira-Spencer theory \cite{BCOV1993}. 

In this work we are concerned with the gravitational corrections to SW theory when it is coupled to ${\cal N}=2$ supergravity background. We review below the calculations of the terms in the topological string theory. We consider two polarizations of the moduli space which in a sense correspond to A and B topological models of string theory compactified on local CY. Then it is shown how the terms are related to SW gravitational corrections. The presentation follows Refs. \cite{Klemm2006,Klemm2009}. The emerging terms, due to their quasi-modular structure, are then related to the topology of a codimension-1 foliations of some compact 3-manifold in the next section. 
 
Let $X$ be a compact CY 3-fold. A complex structure on it is given by the choice of a 3-form $\omega\in H^3(X,\mathbb{C})=H^3(X,\mathbb{Z})\otimes \mathbb{C}$.
Let us consider a symplectic basis $A^I,B_I, I=1,2,...,\frac{b_3}{2}$ of integral homologies $H_3(X,\mathbb{Z})$ of the CY space $X$. Assume, for the simplicity, that the 3-rd Betti number, $b_3(X)=2$. Since $b_3=2(h_{2,1}+1)$ we have $h_{2,1}={\rm dim}_{\mathbb{C}}H^{2,1}(X)=0$. 

The periods serve as the projective parameters (up to a coefficient)
 \[ x=\int_A \omega ,\, p=\int_B \omega .   \]
The moduli space ${\cal M}$ of complex structures on $X$ is a special K\"ahler manifold. 
The relation between periods can be now formulated as the condition for the special geometry on $\cal M$. The holomorphic prepotential ${\cal F}_0$ is determined from the relation between periods: 

\begin{equation}\label{sg}x=\int_A \omega\, ,\;\; p(x)=\frac{\partial}{\partial x}{\cal F}_0(x) . \end{equation}
When topological string theory is on $X$ then ${\cal F}_0$ is given by the classical free energy term of genus zero of that theory. 

For another choice of symplectic basis for $H_3(X,\mathbb{Z})$ and from the equality ${\rm Sp}(2,\mathbb{Z})={\rm SL}(2,\mathbb{Z})$, the new parametric periods are computed as $\tilde{p}=ap+bx, \; \tilde{x}=cp+dx$ which means that $\left(\begin{array}{cc}
a & b \\ c & d \end{array}  \right)\in {\rm SL}(2,\mathbb{Z})$ and $\tau=\frac{\partial}{\partial x}p$ transforms as $\tilde{\tau}=(a\tau +b)(c\tau +d)^{-1}$. Choosing a discrete subgroup $\Gamma \subset {\rm SL}(2,\mathbb{Z})$ one considers complex structures on $X$ which are $\Gamma$-invariant. Since the Teichm\"uller space for the complex structures of the  torus is the upper complex half plane $H^+$, the moduli space of such complex structures is ${\cal M}=H^+/{\Gamma}$. Note the similarity with the moduli space of the SW curve as in \ref{app} for $\Gamma=\Gamma(2),\Gamma(4)$.  

Consider $H^3(X,\mathbb{R})$ as a classical phase space with the symplectic form $\int_X \alpha \wedge \beta$. Every complex structure on $X$ determines one on $H^3(X,\mathbb{R})$ and this complexified space is to be quantized. One has the symplectic form on $H^3(X,\mathbb{Z})$ as ${\rm d}x\wedge {\rm d}p$ and the special geometry condition (\ref{sg}) on CY defines a Lagrangian. The point is that, as explained by Witten \cite{Witten1993}, quantizing the symplectic manifold $H^3(X,\mathbb{Z})$ with respect to the varying complex structures on $X$, the B-model string theory determines a state $|Z\rangle$ in the resulting Hilbert space, such that the topological string partition function can now be written as \cite{Witten1993}:
\begin{equation}\label{Z}
\langle x|Z\rangle=Z(x)=g_s^{\frac{\chi}{24}-1}e^{\sum_{g=0}^{\infty}g_s^{2g-2}{\cal F}_g(x)}\, .
\end{equation} 
Here $\chi$ is the Euler number of $X$,  $g_s$ is the string coupling, ${\cal F}_g$ are the genus $g$ free energy functions of the topological strings and are rather sections of the powers of the prequantum line bundle, than functions. In that way the above expression defines the wave function in the Hilbert space which quantizes the symplectic space $H^3(X,\mathbb{C})$. This Hilbert space is in fact the result of the geometric quantization of $H^3(X,\mathbb{C})$. Moreover, in the quantum theory it holds $[p,x]=g_s^2$ which means that under identification $g_s^2\simeq h/2\pi$ $x,p$ become canonically conjugate operators.

Now let us turn to another polarization of $H^3(X,\mathbb{C})$, namely based on the Hodge decomposition of $H^3$, i.e. $H^3=H^{3,0}\oplus H^{2,1}\oplus H^{1,2}\oplus H^{0,3}$. Still $X$ is compact CY 3-fold but now allow for more general values of $b_3(X)=2n$. Let $\Omega \in H^{3,0}$ be the unique $(3,0)$ form. Given $\omega \in H^3(X,\mathbb{C})$ it can be decomposed as: \[ \omega =\phi \Omega + z D\Omega +\overline{z} \overline{D}\,\overline{\Omega}+\overline{\phi}\overline{\Omega} \]
where as usual, given the K\"ahler potential $K={\rm log}(i\int_X \Omega \wedge \overline{\Omega})$ and $D_i=\partial_i - \partial_i K$, $H^{2,1}$ is spanned by $D_i\Omega$. $(\phi,z^i)$ and their conjugates, become coordinates in the new polarization of the phase space $H^3(X,\mathbb{C})$ and the wave function representing the partition function of the topological string theory, becomes: $\langle z^i,\phi|Z\rangle=Z(z^i,\phi)$. 

Given the symplectic basis again as $(A^I,B_I), I=1,2,...,n$ and introducing $z^I=\phi X^I + z^i D_iX^I$, where $X^I=\int_{A^I} \Omega, P_I=\int_{B_I}\Omega$, the special geometry relation (\ref{sg}) gives again the parameterized periods as $x^I=\int_{A^I} \omega = z^I+c.c$ and $p_I=\int_{B_I} \omega =\tau_{IJ} z^J +c.c$ and $\tau_{IJ}=\frac{\partial}{\partial X^I}P_J$ is now the coupling matrix. The relation between old and new polarization coordinates is given by ${\rm d}p_I\wedge {\rm d}x^I=(\tau_{IJ} - \overline{\tau}_{IJ}){\rm d}z^I\wedge {\rm d} \overline{z}^J$, which can be rewritten in terms of generating function $S$, as ${\rm d}S(x,z)=p_I{\rm d}x^I + (\tau_{IJ} - \overline{\tau}_{IJ})\overline{z}^I{\rm d}z^J$, so that $S(x,z)=\frac{1}{2}\overline{\tau}_{IJ}x^Ix^J + x^Iz^J(\tau_{IJ} - \overline{\tau}_{IJ})+ \frac{1}{2}z^Iz^J(\tau_{IJ} - \overline{\tau}_{IJ})+ const.$. 

When turning to the quantum theory the partition function $\hat{Z}$ in the above (holomorphic) polarization, in terms of $S$, is given by 
\begin{equation}\label{Z-2}
\hat{Z}(z,t,\overline{t})=\int {\rm d}xe^{-S(x,z)/g^2_s}Z(x)\, .
\end{equation}
Here $Z(x)$ is the string partition function for general $n$ \cite{Klemm2006} (which  for $n=1$ is precisely (\ref{Z})) and $t$ is the local coordinate on the moduli space such that $X=X(t)$.

The integral (\ref{Z-2}) can be now expanded around a classical solution described by the moduli space point $x^I_{cl}=X^I$, serving as the saddle point, where $\phi=0, z^i=0$ and $z^I=X^I$. When performing summation over Feynman diagrams where $(-(\tau (X)-\overline{\tau}(\overline{X}))^{-1})^{IJ}$ is the propagator and $\partial_{I_1}...\partial_{I_n}{\cal F}_g(X)$ are the vertices, the result can be written as:
\begin{widetext}
\begin{equation}\label{F-1}
\hat{{\cal F}}_g(t,\overline{t})={\cal F}_g(X)+\Gamma_g((-(\tau (X)-\overline{\tau}(\overline{X}))^{-1})^{IJ}, \partial_{I_1}...\partial_{I_n}{\cal F}_{r<g}(X))\, .
\end{equation}
\end{widetext}
The precise shape of the functionals $\Gamma_g$ can be determined from the path integral by the Feynman rules and expresses the degeneracies of the decomposition of the Riemann surface of genus $g$ into the lower genera curves with the contact terms given by the propagator \cite{Klemm2006}.

Turning to the $n=1$ case the result reads:
\begin{equation}\label{F-2}
\hat{{\cal F}}_g(t,\overline{t})={\cal F}_g(X)+\Gamma_g(-(\tau (X)-\overline{\tau}(\overline{X}))^{-1}, \partial {\cal F}_{r<g}(X))\, .
\end{equation}

Now let us consider the local CY 3-folds still for $n=1$. This enables one to find out the  connection with 4-d $SU(2)$ SW theory in a direct way. Namely, local CY, $X$, is given by the equation in $\mathbb{C}^4$, i.e. $uw=H(y,z)$, which means that CY is the total space of the $\mathbb{C}^{\star}$-fibration over the plane with $y,z$ coordinates and fibers are given by $uw=const$. 
The holomorphic 3-form on $X$ is thus given by $\omega = \frac{{\rm d}u}{u}\wedge {\rm d}y \wedge {\rm d}z$. 

The Riemann surface $\Sigma$ with genus 1 is now determined from $X$ by $H(y,z)=0$ and the 1-form $\lambda=y{\rm d}z$ relates the periods on $X$ with those on $\Sigma$, as:
\begin{equation}\label{form} \int_{A^{(3)}}\omega = \int_{A^{(1)}}\lambda=x,\; \int_{B^{(3)}}\omega = \int_{B^{(1)}}\lambda= p \end{equation} and the periods $x, p$  alone with the additional non-compact periods $\int_C \lambda^i$ parametrize the moduli space. These last, however, as was shown in Ref. \cite{Klemm2006}, do not change the expressions $\hat{{\cal F}}_g(t,\overline{t})$ in (\ref{F-1}) and their dependence on ${\cal F}_{r<g}$ and the functional $\Gamma_g$ in the compact case.    

Conversely, starting from $SU(n)$ SW theory without matter, one determines the shape of $H(y,z)$ from the theory, as: \[ H(y,z)=y^2-\left(z^n+u_2z^{n-2}+... u_n\right)^2 +\Lambda^4 \] and thus, the holomorphic 3-form reads again $\omega = \frac{{\rm d}u}{u}\wedge {\rm d}y \wedge {\rm d}z$. In our case of $SU(2)$, $ H(y,z)=y^2-\left(z^2+u\right)^2 +\Lambda^4 $. The family of Riemann surfaces $y^2-\left(z^2+u\right)^2 +\Lambda^4=0$ is thus, one form of the SW curve (see also \ref{app}). The meromorphic differential $\lambda$ of the SW theory is the reduction of the holomorphic 3-form on CY as in (\ref{form}).  Next, the geometric quantization of the symplectic space $H^3(X,\mathbb{Z})$ results in the geometric quantization of $H_1(\Sigma ,\mathbb{Z})$, and the SW partition function, $Z_{SW}$, is now a wave function in the corresponding Hilbert space (see (\ref{F}) and the discussion after it).

Let us consider now a non-trivial action of a discrete subgroup $\Gamma\subset {\rm SL}(2,\mathbb{Z})$ on the symplectic base of the 3-rd homologies $H_3(X,\mathbb{C})$. Requiring that $\Gamma$ is the symmetry group of the physical string theory means that the topological string partition function and the state $|Z\rangle$ are to be invariant under monodromies generating $\Gamma$. However, the wave function in the 'real' polarization depends on the choice of the symplectic basis in $H_3(X,\mathbb{C})$ which simply means that the wave function is not monodromy invariant.   

On the other hand the wave function in the 'holomorphic' polarization does not depend on any choice of the symplectic basis hence is $\Gamma$-invariant. Looking at (\ref{F-2}) we see that the left hand side is $\Gamma$-invariant (the holomorphic polarization) while the elements of the right hand side are not. For example the term $(\tau - \overline{\tau })^{-1}$ transforms as 
\begin{equation}\label{tau}
(\tau - \overline{\tau })^{-1}\to (c\tau +d)^2(\tau - \overline{\tau })^{-1}-c(c\tau +d)
\end{equation}
where $\left(\begin{array}{cc}
a & b \\ c & d \end{array}  \right)\in \Gamma$. The above non-modularity has to cancel the others from remaining terms of the RHS of (\ref{F-1}), i.e. from ${\cal F}_g$. However, $\hat{{\cal F}}_g$ being modular are not holomorphic. The above relation between $\hat{{\cal F}}_g$ and ${\cal F}_g$ can be summarized as follows \cite{Klemm2006}.  
\begin{itemize}
\item[1.] In the holomorphic polarization, $\hat{{\cal F}}_g(\tau,\overline{\tau})$ is a modular form of $\Gamma$ (of weight 0) which depends on $\overline{\tau}$ through the finite power series in $(\tau - \overline{\tau })^{-1}$.
\item[2.] In the real polarization, the string partition function (\ref{Z}) is determined via ${\cal F}_g(\tau)$ which are holomorphic but only quasi-modular, and ${\cal F}_g(\tau)$ equals to the constant part of the above series of $\hat{{\cal F}}_g(\tau,\overline{\tau})$ expanded in $(\tau - \overline{\tau })^{-1}$. 
\end{itemize}

On the other hand given a discrete subgroup $\Gamma \subset {\rm SL}(2,\mathbb{Z})$ we have the generator of the ring of the quasi-modular forms of $\Gamma$ which is, up to the normalization, the 2-nd Eisenstein series $E_2$ and we denote it by the same symbol $E_2$. It has the following transformation properties 
\begin{equation}
E_2^{\star}(\tau , \overline{\tau })=E_2(\tau )+(\tau -\overline{\tau})^{-1}\, .
\end{equation} 
This means that $E_2^{\star}$ is modular though non-holomorphic. $(\tau -\overline{\tau})^{-1}$ transforms as in (\ref{tau}) so that $E_2(\tau )$ has to transform as: 
\[ E_2(\tau )\to (c\tau +d)^2E_2(\tau ) + c(c\tau + d) \, , \]
and consequently: $E_2^{\star}(\tau , \overline{\tau })\to (c\tau + d)^2 E_2^{\star}(\tau , \overline{\tau })$.

As the result, based on the points $1,2$ above, the free energy at genus $g$ in the holomorphic polarization, read:
\begin{widetext}
\begin{equation}\label{Fg}
\hat{{\cal F}}_g(\tau , \overline{\tau})=h_g^{(0)}(\tau)+h_g^{(1)}(\tau)E_2^{\star}(\tau , \overline{\tau})+...+h_g^{(3g-3)}(\tau)E_2^{\star \, (3g-3)}(\tau , \overline{\tau})\, .
\end{equation}
\end{widetext}
Here all $h_g^{(k)}, k=0,1,...,3g-3$ are holomorphic modular forms of $\Gamma$. Note that taking the $\overline{\tau}\to \infty$ all the terms with $(\tau - \overline{\tau })^{-1}$ decouple and the constant term of the series remains. From points $1,2$ it follows that the real polarization for the free energy functions read: ${\cal F}_g(\tau)={\rm lim}_{\overline{\tau}\to \infty}\hat{{\cal F}}_g(\tau , \overline{\tau})$, so that:
\[{\cal F}_g(\tau)= h_g^{(0)}(\tau)+h_g^{(1)}(\tau)E_2(\tau)+...+h_g^{(3g-3)}(\tau)E_2^{(3g-3)}(\tau) \,. \] 
One of the very important results of Ref. \cite{BCOV1993} was the determination of the equations which showed the non-holomorphic behavior of the amplitudes of the Kodaira-Spencer theory of gravity, hence the topological string amplitudes coupled to the target gravity. These are celebrated holomorphic anomaly equations. The KS theory is realized first by the topological twist of 2-d ${\cal N}=2$ superconformal theory leading to non-gravitational topological QFT, and then, coupling it to gravity. 
Then, the effects at various values of genus $g$ of Riemann surfaces are such that the partition function, ${\cal F}_{g}$, at genus $g\ge 2$ is given by \cite{BCOV1993}:
\[{\cal F}_g=\int_{{\cal M}_g}\langle \prod_{k=1}^{3g-3}(\int_{\Sigma_k} G^-\mu_k)(\int_{\Sigma_k} {\overline{G}^-\overline{\mu}_k})\rangle  \] where ${\cal M}_g$ is the moduli space of the Riemann surface of genus $g$ and $\mu_k$ is the Beltrami differential operator on the surface and $\overline{G}^-$ represents the supercharge of ${\cal N}=2$ 2-d superalgebra (one of the two conjugated right-movers) while $G^-$ represents one of the left movers. As such the dependence on metric is thus written in ${\mu}_k$ making the theory \emph{gravitational}. Such gravitational theory is the topological string theory \cite{Klemm2006,Klemm2007b,Klemm2009}. The very important thing is the possibility to calculate the functions relevant to SW theory from the holomorphic anomaly. Namely, when SW theory on $\mathbb{R}^4$ is coupled to ${\cal N}=2$ supergravity then, the SW Lagrangian (\ref{A-3}) gets the corrections which are encoded in the couplings ${\cal F}_{g}$'s. These couplings appear in terms like:
\begin{equation}\label{F}
{\cal F}_{g}F_+^{2g-2}R^2_+  
\end{equation}
which describe the interactions, and $F_+$ is the self-dual part of the graviphoton field strength and $R_+$ Ricci self-dual curvature. These terms can be grouped in the SW partition function as $Z_{SW}=e^{\lambda^{2g-2}{\cal F}_{g}(a)}$. 

Again, from the point of view of pure $SU(2)$ SW theory (see the \ref{app}) one solves the theory (derives the defining functions of the theory) directly from the modular data. The complex SW family of 1-curves (the Riemann surfaces) can be represented by the equation: \[y^2=(x^2-\Lambda)(x-u) \] and it has 3 singular points which corresponds to the values $u=\pm \Lambda, \infty$. The following monodromies appear:
\[ \begin{array}{cc} M_{\infty}= \left(\begin{array}{cc}
-1 & 2 \\ 0 & -1 \end{array}  \right)\, , \; \; M_{1}= \left(\begin{array}{cc}
1 & 0 \\ -2 & 1 \end{array} \right)\, ,\\[5pt] M_{-1}= \left(\begin{array}{cc}
-1 & 2 \\ -2 & 3 \end{array} \right) . \end{array}\] 
The prepotential ${\cal F}={\cal F}^{pert}+{\cal F}^{inst}$ with its nonperturbative, instantons corrections is given in (\ref{A-5}). It can be calculated from the geometry of the parameter space of the above curves.
Namely, the $u$-plane as the parameter space for elliptic curves, is a special K\"ahler manifold, and the prepotential ${\cal F}$ is included in its definition. For example, the K\"ahler metric on the $u$-plane is the imaginary part of the second derivative of the prepotential,  ${\rm Im}\frac{\partial^2({\cal F})}{\partial a^2}$ (see \ref{app}).
The nonperturbative corrections to the prepotential are the multi-instanton contributions in the weak coupling region and can be calculated by solving the system of the Picard-Fuchs equations. In every of the 3 regions in the quantum moduli space determined from the singularities, one has corresponding expressions for the prepotential  \cite{Lerche1996}, similarly as (\ref{A-5}) is for $a$-coordinates. The gluing of gives the global expressions which are rather sections of some line bundles and modularity is important here. 

The monodromies of the singularities are essential for determining further gravitational corrections to ${\cal F}$, as is already clear from the discussion of the topological string theory side. Again, these are rather quasi-modular forms of the group generated by the monodromies, which govern the corrections. 

In SW theory the corrections can be determined also via direct instanton calculus by Nekrasov. First, the instanton contributions in the prepotential are computed by Nekrasov partition function $Z(\epsilon_1,\epsilon_2,a,a_D)$ which is parameterized by two parameters $\epsilon_1$ and $\epsilon_2$ which deform the $\mathbb{R}^4$ space \cite{Nekr2002}. Then, the leading order contribution of Nekrasov function in small $\epsilon_1$, $\epsilon_2$ is equal to the Seiberg-Witten prepotential ${\cal F}$ \cite{Nekr2003}. The free energy function reads \cite{Min2012}:
\[ {\rm log}\, Z(\epsilon_1,\epsilon_2,a,a_D)=\sum_{g,n=0}^{\infty}(\epsilon_1+\epsilon_2)^{2n}(\epsilon_1 \epsilon_2)^{g-1}F^{n,g}(a,a_D).  \]
Then, $F^{0,0}={\cal F}$ is the prepotential. 
The higher order contributions in $\epsilon_1$, $\epsilon_2$ expansion of the Nekrasov function compute the $n$-instantons contributions to the gravitational coupling terms ${\cal F}_{g > 1}$, as appearing in the effective action (\ref{F}).

Following Ref. \cite{Klemm2009}, let us now consider the holomorphic anomaly equations of the topological string theory, as:
\begin{widetext}
\begin{equation}\label{Anom}
\partial_a \partial_{\overline{a}}{\cal F}_1=\frac{1}{2}C_{aaa}C^{aa}_{\overline{a}}, \; \; \overline{\partial}_{\overline{a}}{\cal F}_g=\frac{1}{2}C^{aa}_{\overline{a}}\left(D_aD_a{\cal F}_{g-1}+\sum_{h=1}^{g-1}D_a{\cal F}_{g-h}{\cal F}_h\right), \, g>1.  
\end{equation}
\end{widetext}
where $C_{aaa}=\frac{\partial^3{\cal F}_0}{\partial a^3}$ and the covariant derivations $D_a=\partial_a - \Gamma_{aa}^a$ are built with respect to the symbols $\Gamma^a_{aa}=(G_{a\overline{a}}^a)^{-1}\partial _a(G_{a\overline{a}}^a)$ and $G_{a\overline{a}}^a$ is the metric on ${\cal M}_g$ - the moduli space of the Riemann surface of the genus $g$, which reads (cf. \ref{app}):
\[ G_{a\overline{a}}^a=8\pi {\rm Im}(\tau)=\frac{\tau -\overline{\tau}}{2i},\; \; {\rm where}\; \; \tau=\frac{1}{4\pi i}\frac{\partial^2 {\cal F}_0}{\partial a^2}=\frac{\Theta}{\pi}+\frac{8\pi i}{g^2}.   \]
Note the appearance of the complex coupling constant $\tau$ for the SW theory which in the \ref{app} was presented in the slightly different normalization. 

The ${\cal F}_1$ function reads: ${\cal F}_1=-\frac{1}{2}{\rm log}({\rm Im}(\tau))-{\rm log}(\eta (\tau))$ \cite{Klemm2006}. Considering (\ref{Anom}) as the recursion in $g$ for ${\cal F}_g$ which encodes the quasi-modularity of the expressions, one shows that the anholomorphic behavior of ${\cal F}_g(\tau,\overline{\tau})$ is entirely by their dependence on $E^{\star}_2$ which gives dual quasi-modular dependence on $E_2$.   
In particular explicit expressions for the leading term of ${\cal F}_g$ in SW theory (without matter multiplet) read \cite{Klemm2009}:
\begin{widetext}
\begin{equation}
{\cal F}_g=\frac{A^g}{(g-1)(1152\cdot 8)^{2(g-1)}}C_{aaa}^{2(g-1)}(E^{\star}_2)^{3(g-1)}+... \; {\rm and}\; A^2=\frac{5}{36},A^3=\frac{5}{18}, A^4=\frac{1105}{1296}...
\end{equation} 
\end{widetext}
Where we can express $C_{aaa}$ as $C_{aaa}=\frac{8\theta_2^2}{\theta_3^4 \theta^4_4}$ and $\theta_k$ are the standard Jacobi $\theta$-functions.

Note that switching between the non-holomorphic expressions, $\hat{{\cal F}}_g, g\ge 2$, and the quasi-modular ones, i.e. ${\cal F}_g$, is achieved simply by the replacing the $E_2^{\star}$ by $E_2$ in the expansions. 

Moreover, one derives ${\cal F}_g$ from the holomorphic anomaly (\ref{Anom}) also for SW theory with $N_f=1,2,3$ massless multiplets. The result is again that $\hat{{\cal F}}_g$ are power series of quasi-modular Eisenstein 2-nd series $E_2$ \cite{Klemm2009}. This ends our quick tour through the quasi-modular structure of the gravitational corrections in SW theory and the connection with the amplitudes of the topological string theory.

\section{GV-twisted modular forms}\label{GV}
There is another way leading to deformed quasi-modular forms from modular ones and which seems to be unrelated with the topological string theory. Namely, as was shown in the first part of this work, exotic smooth $\mathbb{R}^4$'s from the fixed radial family act naturally on the crossed product (see, e.g. Ref. \cite{Sierakowski}) ${\cal M}\ltimes {\rm GL}^+(2,\mathbb{Q})$. The action is determined by the non-trivial GV invariant of the assigned codimension-one foliations. This is in fact the Connes-Moscovici construction of the action of the Hopf algebra representing the universal GV invariant via its cyclic cocycle.

The universal GV class of a codimension one foliation $[\delta_1]$ is represented by the Hopf cyclic cocycle which gives rise to the class in periodic Hopf cohomologies $PHC^1({\cal H}_1)$ of the Hopf algebra ${\cal H}_1$ \cite{ConnesMosc2004,AsselmKrol2012b}. Moreover, given the crossed product ${\cal M}\ltimes {\rm GL}^+(2,\mathbb{Q})$ of the arbitrary level modular forms ${\cal M}$ by the action of the group ${\rm GL}^+(2,\mathbb{Q})$, there is a natural and unique action of $[\delta_1]$ on it. Let us represent the crossed product by the formal finite sums of symbols (spanning the algebra of functions $C({\rm GL}^+(2,\mathbb{Q}),{\cal M})$, see, e.g. Ref. \cite{Sierakowski,Phillips2007}):
\[\sum_{\gamma} fU^{\star}_\gamma \, ,\;\; {\rm where}\; f\in {\cal M}\; {\rm and}\;  \gamma \in {\rm GL}^+(2,\mathbb{Q}) \] where $U_{\gamma}^{\star}$ corresponds to the automorphism of ${\cal M}$, and the product reads:
\[fU^{\star}_{\alpha} \cdot g U^{\star}_{\beta} =(f\cdot g|\alpha)U^{\star}_{\beta \alpha} \] where 
\begin{equation*}\label{1}
\begin{array}{c}
g|_{k}\alpha:=\sqrt{\det{\alpha}}g\left(\frac{a_1z+a_2}{a_3z+a_4}\right)(a_3z+a_4)^{-k} \\[5pt]
{\rm for}\; \; \alpha = \left(\begin{array}{cc}
a_1 & a_2\\
a_3 & a_4\end{array} \right)
 \in {\rm GL}^+(2,\mathbb{Q}) \,. 
\end{array}
\end{equation*} 

Then, the action of $[\delta_1]$ on $fU_{\gamma}^{\star}$ is given by 
\begin{equation}\label{Mod-2}
\delta_1(fU^{\star}_{\gamma})=\mu_{\gamma}\cdot fU^{\star}_{\gamma}
\end{equation}
where $\mu_{\gamma}$ is a (normalized) non-holomorphic modular Eisenstein series of weight 2 for every $\gamma = \left(\begin{array}{cc}
a & b\\
c & d\end{array} \right) \in {\rm GL}^+(2,\mathbb{Q})$ which reads:
\[\mu_{\gamma}(z)=\frac{1}{2\pi^2}\left(G_2^{\star}|\gamma(z)-G_2^{\star}(z)+\frac{2\pi ic}{cz+d}\right)\,.   \] $G^{\star}_2=\frac{\pi^2}{3}-8\pi^2\sum_{m,n\leq 1}me^{2\pi imnz}$ is the (up to some normalization factor) holomorphic Eisenstein series of weight 2 which fails to be modular, i.e.:
\[G^{\star}_2|\alpha (z)=G^{\star}_2(z)-\frac{2\pi ic}{cz+d},\; \alpha =  \left(\begin{array}{cc}
a & b\\
c & d\end{array} \right)\in \Gamma (1)\,. \]
One can rewrite $\mu_{\gamma}$ in terms of $E_2(z)$ the generator of the non-modular 2-nd Eisenstein series which were already in the extensive use in the previous section. Namely, \[ G_2(z)=\zeta(2)E_2(z)\] where $\zeta(2)=\sum_{r\geq 1}\frac{1}{r^2}$ is the value at 2 of the Riemann zeta function. Thus, \begin{equation}\label{Mod-3} \mu_{\gamma}(z)= \zeta(2)E_2(z) \end{equation} and the action (\ref{Mod-2}) reads:
\begin{equation}\label{Mod-4} \delta_1(fU^{\star}_{\gamma})=\zeta(2)E_2(z)\cdot fU^{\star}_{\gamma}. \end{equation}
Let us note that the 2-nd Eisenstein series $E_2$ generates the ring of quasi modular forms which is isomorphic to the dual ring of almost holomorphic functions generated by $E_2^{\star}$. Given the definition of the general modular Eisenstein series for $k> 2$: \[ G_k(z)=\frac{1}{2}\sum_{m,n\in \mathbb{Z}\setminus \{(0,0)\}}\frac{1}{(mz+n)^k} \] and the relation \[ G_k(z)=\zeta(k)E_2(z),\] the following proposition explains the role of $E_4$ and $E_8$:
\begin{lemma} (Prop. 4, Ref. \cite{Zagier2008})
The ring $M_{\star}{\Gamma(1)}$ of modular forms of every weight is freely generated by the modular forms $E_4$ and $E_8$.
\end{lemma}
For other subgroups $\Gamma(N)$ of $\Gamma(1)=\rm{SL}(2,\mathbb{Z})$ their Eisenstein quasi-modular series are also generated by $E_2$ \cite{Zagier2008} and the action (\ref{Mod-2}) is still valid for $\mu_{\gamma}$ as in (\ref{Mod-3}). 

Suppose that certain terms of the Lagrangian of some QFT coupled to a gravity background appear, in the topological limit of QFT, as the functions of the power series in $E_2$ with modular coefficients, i.e. $\sum_{i=0}^{n}f_i\cdot (E_2)^i,f_i\in {\cal M}$ or, the correlation functions of such theory acquire such a structure. Suppose also that in another region of the moduli space of the theory, the above expressions become corresponding functions of the non-holomorphic $E_2^{\star}$.  
Then, we say that the theory \emph{respects the topology of a codimension-one foliation} of the background in this limit. In the remaining part of this paper we will show that such situation indeed can be substantiated by some gauge theory, string theory and a background 4-geometry. 

\section{Gravitational corrections to Seiberg-Witten theory and exotic $\mathbb{R}^4$}\label{Grav-Corr}
We know that exotic $\mathbb{R}^4$'s are non-flat Riemannian smooth 4-manifolds which probably support nontrivial gravitaional backgrounds \cite{Sladkowski2001,Asselm-Krol-2011}. Let us consider metrics on exotic $\mathbb{R}^4$'s from the radial family as a dynamical gravitational variables. Moreover, we are interested in exotic 4-geometry which emerges from some QG description. This point of view was followed recently from different perspective  (see, e.g. Refs. \cite{AsselmKrol2011d,AsselmeyerKrol2011,QG-2012,Krol2011d,Krol2010}) where superstring theory was taken as a working theory of QG. We want to derive further conclusions from such general scenario. Such thinking appears indeed as potentially fruitful for both, superstring theory where new 4-d ingredients emerge, and exotic $\mathbb{R}^4$ where mathematics of superstrings is applicable \cite{AssKrol2010ICM,AsselmKrol2011d,AsselmeyerKrol2011,AsselmeyerKrol2011b,QG-2012}.

Let us be more specific about a theory of gravity relevant here and about YM theory on exotic $\mathbb{R}^4$. Supersymmetric ${\cal N}=2$ YM theory (on flat $\mathbb{R}^4$) has 8 conserved supercharges, $Q^i_{\alpha}, Q^i_{\dot{\alpha}}$ which correspond to the global symmetry group $SU(2)_L\times SU(2)_R\times SU(2)_I$. First two $SU(2)$-factors are due to the group of spatial rotations of 4-space, the last one is the $R$-symmetry and the indices $\alpha, \dot{\alpha}, i$ are dublets of the corresponding factors. 
When formulating supersymmetric ${\cal N}=2$ YM theory on a smooth 4-manifold the amount of preserved supercharges can be, in general, reduced. Still, at least 1 supercharge is always preserved on every smooth 4-manifold\footnote{We thank Edward Witten for explaining this to us.}. Namely, performing the twist in ${\cal N}=2$ YM theory, such that the diagonal subgroup $SU(2)_d$ of $SU(2)_R\times SU(2)_I$ is chosen, one gets modified supercharges $Q,Q^+_{\mu\nu},G_{\mu}$ where now the spacetime indices are with respect to $SU(2)_L\times SU(2)_d$ subgroup. $Q$ is the superccharge which is preserved on every smooth 4-manifold, and usually refers to the BRST charge in the topological quantum field theory twist of supersymmetric YM theory. The self-dual 2-form supercharge $Q^+_{\mu\nu}$ is preserved on every K\"ahler 4-manifold and was used by Witten, e.g. Ref. \cite{Witten1994}. The remaining $G_{\mu}$ supercharges became essential in the equivariant reformulation of the theory and in the calculation of instantons corrections by Nekrasov \cite{Nekr2002}. 

The full amount of 8 supercharges cannot be preserved when ${\cal N}=2$ YM is formulated on exotic $\mathbb{R}^4$, let alone the requirement of translational invariance causes the smooth $\mathbb{R}^4$ becomes the standard one. Still, at least 1 supercharge can be preserved as discussed above. Precisely this supercharge $Q$ was responsible for deriving the observables of the twisted SYM theory, namely those which are annihilated by $\overline{Q}$. The correlation functions of these observables are metric independent and become the smooth structure invariants of the 4-manifold. In the case of some compact smooth 4-manifolds the invariants calculated from $\overline{Q}$ and ${\rm Tr}\Phi^2$ (the component of the superfield ${\cal N}=2$ vector supermultiplet) are celebrated Donaldson polynomial invariants. Taking abelian $U(1)$ gauge symmetry and with the presence of a single hypermultiplet the correlators become Seiberg-Witten monopole smooth invariants. Seiberg-Witten equations alone can be formulated on every oriented smooth 4-manifold \cite{Taubes1997}. The theory whose correlation functions are the smooth invariants is the twisted SYM topological theory and can be formulated on every 4-manifold, hence on exotic $\mathbb{R}^4$ as well. Even though the case of exotic $\mathbb{R}^4$ is untractable by these means currently (first Betti number $b_+=0$, non-compact), some interesting exotic invariants might be derivable, in principle, as the correlation functions of (twisted) SYM theory with the supercharge $\overline{Q}$ on these manifolds. Instead, in this paper we approach some results on exotic $\mathbb{R}^4$, in the above context, in a way which refers rather to geometry of exotic 4-spaces than makes direct use of supersymmetric gauge theories. We turn to this issue now.

Let us call the theory resulting from the ${\cal N}=2$ YM formulated on exotic $\mathbb{R}^4$, the \emph{SYM theory on exotic $\mathbb{R}^4$}. Also here at least one supercharge say, $\overline{Q}$, is preserved. We always understand that exotic $\mathbb{R}^4$ belongs to the fixed radial family. As was discussed already in Sec. \ref{strings-1} Seiberg-Witten theory is the effective low energy limit of supersymmetric ${\cal N}=2$ Yang-Mills theory and the Lagrangian of that theory was worked out in Refs. \cite{Seiberg1988,Witten-Seiberg1994} (see \ref{app}). Again, this effective theory can be formulated on curved 4-manifolds. Let $\phi$ is the scalar field related by supersymmetry to the gauge field and $u={\rm Tr}{\phi}^2$ is the usual variable in the moduli space of the theory. The effective SW Lagrangian acquires corrections due to the curvature of the background manifold which, in general, are couplings of $u$ to the polynomials of Riemann tensor and its derivatives \cite{Witten1995}.  
Even though exotic $\mathbb{R}^4$ are currently rather untractable analytically, our purpose is to grasp the above mentioned gravitational corrections to SW Lagrangian due to curved exotic $\mathbb{R}^4$, by the structure of exotic 4-spaces.

SW theory as supersymmetric ${\cal N}=2$ YM couples naturally to ${\cal N}=2$ supergravity. Thus, the idea is such that this kind of supergravity and the coupling expresses the gravitational effects of exotic $\mathbb{R}^4$ when YM theory is formulated on that background. 
In fact the coupling of SW to ${\cal N}=2$ supergravity expresses dynamical aspects of metrics on exotic $\mathbb{R}^4$.

From the technical point of view the ${\cal N}=2$ case is crucial since under these supersymmetries 4-d Yang-Mills theory is exactly solvable in low energies and is governed by a single holomorphic prepotential ${\cal F}$, as we already saw in Sec. \ref{strings-1} (cf. Ref. \cite{Lerche1996}). That is why the corrections to SW, due to the exotic open 4-smoothness, can be derived in such set-up effectively as the corrections to a single function - the prepotential of $SU(2)$ SW on flat $\mathbb{R}^4$. This is possible, however, under some further natural assumptions, which we are now going to clarify.

It has been explained in the first part of this work \cite{AsselmKrol2012b} that the members of the fixed radial family determines codimension-1 foliations of some 3-d compact manifolds.
We want to consider a special limit of SYM theory on exotic $\mathbb{R}^4$ which deals with exotic foliated topology underlying the smoothness. Let us define \emph{foliated topological} limit of SYM theory on exotic $\mathbb{R}^4$ as low energy limit of the theory, such that:
\begin{itemize}
\item[i.] In this low energy limit one deals consistently with gravitational corrections to flat SW on $\mathbb{R}^4$;
\item[ii.] The topology of the codimension-1 foliations of some compact 3-manifolds become important and generates dominant corrections to the 'flat' SW  Lagrangian on $\mathbb{R}^4$;
\item[iii.] The corrections are given by some polynomials of ${E}_2$ with modular coefficients.
\end{itemize}
For a given radial family of small exotic smooth $\mathbb{R}^4$'s let $e$ be some member of the family. Let us also suppose that the foliated topological limit is unique in the sense that whenever the gravitational corrections to the Lagrangian in the topological limit are polynomials of ${E}_2$ these are \emph{uniquely} determined by the coupling of SW to ${\cal N}=2$ supergravity. In that way we arrive at the following result: 
\begin{lemma}
The foliated topological limit of $SU(2)$ SYM on $e$ is given by $SU(2)$ SW on standard-flat $\mathbb{R}^4$ coupled to ${\cal N}=2$ supergravity. The same holds true for the $SU(2)$ SW with $N_f=1,2,3$ massless matter multiplets.
\end{lemma} 
{\em Proof.} Let us show how the SW coupled to ${\cal N}=2$ supergravity fulfils the requirements i., ii. and iii. above. It is almost tautological. First, in the topological foliated limit the coupling of SW to gravity on $e$ should be achieved, hence the corresponding corrections to the flat SW Lagrangian emerge. Next, codimension-1 foliations of $S^3$ act on the modular algebra ${\cal M}$ via multiplying by ${E}_2$ according to the Theorem 9 from Ref. \cite{AsselmKrol2012b}. Thus, the polynomials of ${E}_2$ with modular coefficients emerge as the proposed corrections in the limit where contributions from the foliations dominate. However, according to the uniqueness supposition these are ${\cal F}^{(g)}$'s terms of SW coupled to ${\cal N}=2$ supergravity. Thus, when the topological foliated limit on $e$ is achieved it is given by SW on standard-flat $\mathbb{R}^4$ coupled to ${\cal N}=2$ supergravity. The similar polynomials in $E_2$ structure of the gravitational corrections is obtained in the case of $SU(2)$ SW with massless matter multiplets, and these become polynomials in $E^{\star}$ in the holomorphic polarization \cite{Klemm2009}. 

As the another immediate consequence one has: 
\begin{corollary}
The polynomials ${\cal F}^{(g)}$ of $E_2$ are determined in the foliated topological limit of SYM on $e$ as the gravitational corrections to the flat $SU(2)$ SW Lagrangian with $N_f=0,1,2,3$ massless matter multiplets.
\end{corollary} 
Note that the corrections derived from the foliated topological limit does not depend on the choice of one or another exotic $\mathbb{R}^4$ from the radial family. Rather, every member of the radial family gives the same contribution. If one would like to distinguish between different exotic $\mathbb{R}^4$'s the switch from the universal GV class to the contributions depending on the actual values of GV invariant, is needed. The result would be, presumably, different polynomials in ${E}_2$ with modular coefficients than these given by ${\cal F}^{(g)}$'s. This foliated non-topological limit (depending on exotic 4-metrics) will not be, however, considered here since the metrics are not in reach up to present understanding of the subject. However, such expressions would be a step toward generating proper invariants of exotic $\mathbb{R}^4$ from the fixed radial family. Still some relative results would be possible to work out. 
\begin{theorem}\label{SW-string}
The foliated topological limit of SYM on $e$ is given by the topological superstring correlation functions on certain local CY 3-fold.
\end{theorem}
 {\em Proof.}
Again, we have the universal twisting of arbitrary level modular forms by Eisenstein series. This is derived from the radial family as in Ref. \cite{AsselmKrol2012b} and recalled in the previous section. Given the results presented in Sec. \ref{strings-1}, the twisting is seen in the topological string amplitudes $F^{(g)}$ where they are the polynomials in ${E}_2$ and are obtained as the correlation functions of the topological string theory B-model in some local CY manifold. 

\begin{corollary}
Exotic smooth structure on $\mathbb{R}^4$ from the fixed radial family, generates the gravitational corrections which in the foliated topological limit of SYM on $e$ are equal to the corrections to $SU(2)$ SW Lagrangian, which are calculated by the instanton higher  gravitational corrections of Nekrasov.
\end{corollary}
This corollary is in fact the reformulation of the results related with the instanton calculus discussed in Sec. \ref{strings-1}.
  
Th. \ref{SW-string} indicates a possible role of the radial family of exotic $\mathbb{R}^4$'s in the appearance/resolution of the holomorphic anomaly as was also briefly discussed in Sec. \ref{strings-1}.
 
Thus, the general shape of $F^{(g)}$'s show the quasi-modular deformation, which is due to the holomorphic anomaly on the string theory side, whose building blocks are polynomials of ${E}_2$ with modular forms as coefficients. The following proposition is the reformulation of i. and ii. above:

\emph{In the special topological (foliated) limit, $SU(2)$ YM theory on exotic $\mathbb{R}^4$'s from the radial family with dynamical metrics, is given by the supersymmetric $SU(2)$ SW coupled to ${\cal N}=2$ supergravity. The gravitational corrections to the SW Lagrangian are then the polynomials in ${E}_2$, and are the gravitational corrections due to the exotic $\mathbb{R}^4$.}

In the non-topological foliated limit, where metrics on $e$ should matter, the calculations of the superstring correlation functions should again give 'deformed' modular polynomials of $E_2$, i.e. more general functions of $E_2$. 

\section{Discussion and perspectives}
In the Kodaira-Spencer theory of gravity \cite{BCOV1993} the propagator of the theory is precisely ${E}_2$ which is reflected in the polynomial structure of the amplitudes of the string theory B model on local CY, \cite{Klemm2009} which was discussed in Sec. \ref{strings-1}. Let us consider the exotic $\mathbb{R}^4$ embedded in the local 3-fold CY. From the point of view of higher dimensions the exotic 4-geometry is smoothly embedded in six real dimensions. However, in 4-d it is confined to an open subset of the Euclidean flat 4-space, producing singular non-smooth imbedding. Having in mind that the radial family acts on modular forms via ${E}_2$ - the propagator of the gravitational theory, and it gives formally amplitudes of string theory which represent the gravitational correction to the topological theory, the following conjecture is in order:

\emph{The KS theory of gravity describes gravitational corrections to $\sigma$-model in the CY target which are precisely generated by exotic geometry of $\mathbb{R}^4$ chosen from the fixed radial family embedded in this local CY. From the point of view of 4-d these are gravitational corrections to $SU(2)$ SW theory and thus, gravitons in 4-d could correspond to exotic 4-geometry on $\mathbb{R}^4$ in some fundamental 4-d theory of QG.}

In the sense of the above conjecture and in the language of semiclassical approximation to QG, one could try to identify gravitons with exotic geometry on $\mathbb{R}^4$. Semiclassically, graviton is seen as the perturbation of the flat metric on $\mathbb{R}^4$. Locally one has usual approximation, $g_{\mu \nu}=\eta_{\mu \nu}+e_{\mu\nu}, \mu , \nu = 0,1,2,3$, where $\eta_{\mu\nu}$ is the flat Euclidean metric and $e_{\mu\nu}$ its small perturbation. However, in global picture, $e_{\mu\nu}$ are local presentation of some metric on exotic $e$. Global, also non-perurbative, effects become important and are captured by the fake smooth structure of $\mathbb{R}^4$. Thus, the perturbations of the metric are not necessary small which indicates non-locality of such theory of QG. On the other hand, the connection with superstring theory and supersymmetric YM in 4-d is written in the structure of the theory as presented in the paper. Thus, the global QG corrections from exotic $\mathbb{R}^4$ are calculable via topological string and SW theories. This interesting new aspect of a theory of QG in 4-d deserve further study which will be performed in separate papers. 
Moreover, no supersymmetry is, in principle, required by the above semiclassical approximation and the GV twist of modular forms does not make any use of supersymmetry hence, in principle, exotic $\mathbb{R}^4$ from the fixed radial family can give similar modular behavior of 'topological' correlation functions as in the presence of supersymmetry. 

Note that the supersymmetric YM after the topological twist calculates Donaldson polynomials of the compact 4-manifolds as its correlation functions.
On the other hand, twisted topological string and SW correlation functions could grasp some effects of exotic geometry on $\mathbb{R}^4$ (see also Ref. \cite{AsselmKrol2011f}).
Thus, one could wonder whether and up to what extend the effects assigned to supersymmetric YM can appear in non-supersymmetric YM on exotic $\mathbb{R}^4$. 

Gravity confined to exotic 4-geometry couples indeed, at least formally, to non-supersymmetric YM as in, e.g., the semiclassical case discussed above, or when YM theory is formulated on exotic $\mathbb{R}^4$. In some limit, one could deal with a similar quasi-modular corrections as those specific to the supersymmetric ${\cal N}=2$ YM coupled to ${\cal N}=2$ supergravity.
Exotic $\mathbb{R}^4$'s are Euclidean smooth 4-manifolds hence the 'gravitational' corrections they cause are presumably non-perturbative, instanton-like, and precisely these are exactly calculable in the supersymmetric case (see e.g. Ref. \cite{Nekr2002}). The quasi-modularity of the expressions in the non-supersymmetric case is derived from the codimension-one foliations of $S^3$ rather than from susy. Such effects are completely supersymmetry independent. 

An intriguing possibility thus, emerges, namely a kind of continuous shift between supersymmetric YM, ${\cal N}=2$ with gravitational corrections, and non-supersymmetric YM though on exotic $\mathbb{R}^4$. The technical reason behind such correspondence is the fact that gravitational corrections are calculated as derivative of prepotential which defines SW theory.  

\emph{$SU(2)$ YM on exotic $\mathbb{R}^4$'s from the radial family acquires gravitational corrections which, in some limit can be retrieved as the supersymmetric SW gravitational corrections. Possibly, QCD on exotic $\mathbb{R}^4$ can be continuously related to the gravitational corrections to $SU(3)$ SW theory on flat $\mathbb{R}^4$.}

From that point of view exotic 4-geometries on $\mathbb{R}^4$ could be considered as a formal link connecting the confinement and mass gap appearing in supersymmetric SW theory, with non-supersymmetric 4d YM on exotic $\mathbb{R}^4$.

The analysis of an eventual influence of the open 4-exoticness on the mass gap and confinement problems in the realistic YM theories, is worth pursuing and will be performed in the separate work.

\section*{Acknowledgments}
We thank Edward Witten for explaining some points of SW theory on exotic $\mathbb{R}^4$, to us.

\appendix

\section{$SU(2)$ ${\cal N}=2$ SYM and SW prepotential}\label{app}
Here we collect some very basic formulas regarding SYM theory starting from susy manipulations leading to the effective SW theory and its modular properties of the SW curve determining the ${\cal N}=2$ prepotential. Nevertheless, such, rather express presentation, may help understanding some points in the paper, and is based mainly on Refs. \cite{Bilal1996,Lerche1996}.
 
In the case of ${\cal N}=1$ one represents the susy algebra via complex scalar field $\phi$ and 2-component spinor field $(\psi_1,\psi_2)=\psi_a, a=1,2$ which form \emph{chiral scalar multiplet}. $\phi$ and $\psi$ are organized in a chiral superfield on the superspace with fermionic anticommuting variables $\theta^{\alpha},\overline{\theta}_{\dot{\alpha}}$:
\[\Phi=A(y)+\sqrt{2}\theta\psi(y)+\theta \theta F(y)  \] where as usual $y^{\mu}=x^{\mu}+i\theta \sigma^{\mu}\overline{\theta},\theta \psi=\theta^{\alpha}\psi_{\alpha},\theta \theta=\theta^{\alpha} \theta_{\alpha}=-2\theta^1\theta^2,\theta \sigma^{\mu}\overline{\theta}=\theta^{\alpha} \sigma^{\mu}_{\alpha \dot{\alpha}}{\overline{\theta}} ^{\dot\alpha}$ and $F$ is the axillary field.
Then, the superspace integral $ \frac{1}{4}\int {\rm d}^4x{\rm d}^2 \theta \Phi^+\Phi $ gives rise to the supersymmetry invariant action. Performing the $\theta$ integrations the action reads: \[ \int {\rm d}^4x(\partial _{\mu}A\partial ^{\mu}A^+-i\overline{\psi}\overline{\sigma}^{\mu}\partial_{\mu}\psi +F^+F) \] where the standard kinetic terms appear. Introducing susy invariant interactions one uses chiral superpotential ${\cal W}(\Phi)$ via $\int {\rm d}^4x[\int {\rm d}^2\theta {\cal W}(\Phi)+...]$.

Another representation of ${\cal N}=1$ susy is build from the \emph{vector} multiplet, containing massless gauge vector field $A_{\mu}$ and the gaugino $\lambda_{\alpha}$ superpartners. Again, the superfield $V$ on the superspace with the axillary fields $D$ is a useful tool: 
\[V=-\theta\sigma^{\mu}\overline{\theta}A_{\mu}+i\theta \theta (\overline{\theta}\overline{\lambda})-i\overline{\theta}\overline{\theta}(\theta \lambda)+\frac{1}{2}\theta \theta \overline{\theta}\overline{\theta}D.   \] 
All the fields are in the adjoint representation of $SU(2)$. Next, one defines spinorial superfield $W$:
\[ W=(-i\lambda + \theta D- i\sigma^{\mu \nu}\theta F_{\mu\nu}+\theta \theta \sigma^{\mu}\nabla _{\mu}\overline{\lambda}) \] where $F_{\mu\nu}=\partial_{\mu}A_{\nu}-\partial_{\nu}A_{\mu}-ig[A_{\mu},A_{\nu}]$ and $\nabla_{\mu}\lambda=\partial_{\mu}\lambda-ig[A_{\mu},\lambda]$. Introducing usual superspace derivatives, $D_{\alpha}=\frac{\partial}{\partial \theta^{\alpha}}+i\sigma^{\mu}_{\alpha \dot\alpha}\overline{\theta}^{\dot{\alpha}}\partial _{\mu}$, $\overline{D}_{\dot\alpha}=-\frac{\partial}{\partial \overline{\theta}^{\dot\alpha}}-i\sigma^{\mu}_{\alpha \dot\alpha}\theta^{\alpha}\partial _{\mu}$, the supersymmetric action reads 
\begin{widetext}
\[-\frac{1}{4}\int {\rm d}^4x {\rm d}^2\theta {\rm tr}W^{\alpha}W_{\alpha} =\int {\rm d}^4x
 {\rm tr}\left[ -\frac{1}{4}F_{\mu\nu}F^{\mu\nu}+\frac{i}{4}F_{\mu\nu}\tilde{F}^{\mu\nu}-i\lambda\sigma^{\mu}\nabla_{\mu}\overline{\lambda}+ \frac{1}{2}D^2\right]. \]
\end{widetext}
$\frac{i}{4}\int {\rm d}^4x F_{\mu\nu}\tilde{F}^{\mu\nu}$ is the instanton number. Introducing the complex coupling constant
$ \tau= \frac{\Theta}{2\pi}+\frac{4\pi i}{g^2} $ one gets: 
\begin{widetext}
\[ \frac{1}{16\pi}{\rm Im}\left[\tau\int {\rm d}^4x {\rm d}^2\theta {\rm tr}W^{\alpha}W_{\alpha}\right] =\frac{1}{g^2}\int {\rm d}^4x {\rm tr}\left[ -\frac{1}{4}F_{\mu\nu}F^{\mu\nu}-i\lambda\sigma^{\mu}\nabla_{\mu}\overline{\lambda}+ \frac{1}{2}D^2\right] +\frac{\Theta}{32\pi^2}\int {\rm d}^4xF_{\mu\nu}\tilde{F}^{\mu\nu} .  \]
\end{widetext}
Turning to ${\cal N}=2$ susy all the fields $A , \psi , A_{\mu},\lambda $ are in the same susy multiplet and thus, are in the same, adjoint, representation of $SU(2)$. The susy invariant action now reads:
\begin{widetext}
\begin{equation*}
\begin{array}{c}
S={\rm Im}\,{\rm tr}\int {\rm d}^4x\frac{\tau }{16\pi}\left[\int {\rm d}^2\theta W^{\alpha}W_{\alpha}+\int {\rm d}^2\theta {\rm d}^2\overline{\theta}\Phi^+e^{-2gV}\Phi \right]=\\[5pt]
\frac{1}{g^2}\int {\rm d}^4x {\rm tr}\left[ -\frac{1}{4}F_{\mu\nu}F^{\mu\nu}-i\lambda\sigma^{\mu}\nabla_{\mu}\overline{\lambda}+ \frac{1}{2}D^2\right] +\frac{\Theta}{32\pi^2}\int {\rm d}^4xF_{\mu\nu}\tilde{F}^{\mu\nu} +\\[5pt]
\int {\rm d}^4x {\rm tr}\left(|\nabla_{\mu}A^2-i\overline{\psi}\overline{\sigma}^{\mu}\nabla_{\mu}\psi +F^+F-gA^+[D,A]-\sqrt{2}igA^+\{\lambda ,\psi \}+\sqrt{2}ig\overline{\psi}[\overline{\lambda},A]\right) .
\end{array}\label{App-2}
\end{equation*}
\end{widetext}
Grouping the axillary fields $D,F$ as $S_{D,F}=\frac{1}{g^2}\int {\rm d}^4x\, {\rm tr}\left[\frac{1}{2}D^2-gA^+[D,A ]+F^+F\right]$ and solving the field equations for them, results in \[ S_{D,F}=-\int {\rm d}^4x\frac{1}{2}{\rm tr}\, ([A^+,A])^2.    \]
Thus, to preserve susy the bosonic classical potential should vanish, $V(A)=\frac{1}{2}{\rm tr}\, ([A^+,A])^2=0$, which means that in the vacuum configuration $A_0$ and $A_0^+$ commute. 

In the superspace notation, adding additional set of anticommuting coordinates $\tilde{\theta}_{\alpha},\overline{\tilde{\theta}}_{\dot{\alpha}}$, the chiral superfield reads: $\Psi=\Phi(\tilde{y},\theta)+\sqrt{2}\tilde{\theta}^{\alpha}W_{\alpha}(\tilde{y},\theta)+\tilde{\theta}^{\alpha}\tilde{\theta}_{\alpha}(\tilde{y},\theta)$ where now $\tilde{y}^{\mu}=x^{\mu}+i\theta \sigma^{\mu}\overline{\theta}+i\tilde{\theta}\sigma^{\mu}\overline{\tilde{\theta}}$. The ${\cal N}=2$ susy invariant action is given by \[S={\rm Im}\, \left[\frac{\tau}{16\pi}\int {\rm d}^4x{\rm d}^2\theta {\rm d}^2\tilde{\theta}\frac{1}{2}{\rm tr}\, \Psi^2 \right].   \] 
In general the ${\cal N}=2$ susy invariant action reads \[{\rm Im}\, \frac{1}{16\pi}\int {\rm d}^4x{\rm d}^2\theta {\rm d}^2\tilde{\theta}{\cal F}(\Psi)\] where ${\cal F}$ is the 'holomorphic' (does not depend on $\overline{\Psi}$) prepotential. For ${\cal N}=2$  it reads ${\cal F}(\Psi)=\frac{1}{2}{\rm tr}\, \tau \Psi^2$.
This high energy microscopic theory is renormalizable and asymptotically free. The SW theory is the effective low energy approximation of that theory. As was noted above non-zero $A$ is allowed in order to preserve ${\cal N}=2$ susy $SU(2)$ gauge theory. Gauge invariance and $[A , A^+]=0$ gives $A=\frac{1}{2}a\sigma_3$ and semiclassically ${\rm tr}\, A^2=\frac{1}{2}a^2$. Quantum-mechanically the gauge inequivalent vacua of the theory are distinguished by the complex value $u=\langle {\rm tr}\, A^2 \rangle$. Thus, the moduli space of the theory, ${\cal M}$, is the complex $u$-plane. The important thing is that the behavior near the singularities of ${\cal M}$ determines the low energy effective theory. $SU(2)$ gauge symmetry is broken to the $U(1)$ by non-zero $\langle \phi \rangle $, though ${\cal N}=2$ susy is preserved. Here $\phi$ is the scalar component of the ${\cal N}=2$ vector multiplet $(A_{\mu},\lambda,\psi,\phi)$. $\phi$ while non-zero gives masses, via Higgs mechanism, to gauge fields breaking $SU(2)$. Thus, the low energy Lagrangian in terms of the massless ${\cal N}=1$ chiral $A$ and vector superfields $W_{\alpha}$, where $SU(2)$ is broken to $U(1)$, reads:
\begin{equation}\label{A-3} {\cal L}_{eff}=\frac{1}{4\pi}{\rm Im}\int {\rm d}^4x\left[\int d^4\theta \frac{\partial {\cal F}}{\partial A}A^+ +\frac{1}{2}\int d^2\theta \frac{\partial ^2{\cal F}}{\partial ^2A}W_{\alpha}W^{\alpha} \right]. \end{equation} 
From the component-like presentation of the above effective action one sees that 4-d $\sigma$-model description emerges and ${\rm Im}\,\frac{\partial^2 {\cal F}}{\partial A^2}$ is a metric on the space of fields, and thus, ${\rm Im}\,\frac{\partial^2 {\cal F}}{\partial a^2}$ is the metric on the moduli space ${\cal M}$. The effective complex coupling constant reads $\tau (a)=\frac{\partial^2 {\cal F}}{\partial a^2}$.  However, there are singular points, as e.g. ${\rm Im}\, \tau(a) \to 0$, and one uses the dual coordinates description of ${\cal M}$, i.e. $a_D$. The singularity is merely coordinate-like and the $\tau(a)$ is rather a section of a bundle than a function. The bundle is in fact ${\rm SL(2,\mathbb{Z})}$-bundle on ${\cal M}$. However, the structure of singularities at quantum theory determines the prepotential ${\cal F}$ and its non-perturbative instantons corrections. Namely, one has 3 singular points in ${\cal M}$ which correspond to $u=\pm \Lambda^2$ and $u=+\infty$ ($\Lambda$ being the scale of the effective theory). The three regions of ${\cal M}$ emerges with three distinct weakly coupled (perturbative) descriptions of the theory around the singular points by local effective Lagrangians. When one glues together these descriptions the nontrivial monodromies emerge by loops circling the singularities:
\[ \begin{array}{cc} M_{\infty}= \left(\begin{array}{cc}
-1 & 4 \\ 0 & -1 \end{array}  \right)\, , \; \; M_{+\Lambda^2}= \left(\begin{array}{cc}
1 & 0 \\ -2 & 1 \end{array} \right) \, , \\[5pt] M_{-\Lambda^2}= \left(\begin{array}{cc}
-1 & 4 \\ -1 & 3 \end{array} \right)\, .\end{array}\] 
For ${\cal N}=2$ superalgebra its central charge is expressed via the magnetic $g$ and electric $q$ quantum numbers of the BPS state of SYM theory 
$Z=qa+ga_D  $ and the dual variable is given by $a_D=\frac{\partial {\cal F}(a)}{\partial a}$. Thus, the global consistency conditions of the patching together can be also expressed as $M_{+\Lambda^2}\cdot M_{-\Lambda^2}=M_{\infty}$ where $M_{+\Lambda^2}=M^{(0,1)}$, $M_{-\Lambda^2}=M^{(1,-2)}$ and $M^{(g,q)}=\left(\begin{array}{cc}
1+qg & q^2 \\ -g^2 & 1-gq \end{array}  \right)$.

Now the reverse problem arises namely, how to determine the functions $a(u),a_D(u)$ with the above monodromies around the singularities which also gives ${\rm Im}\,\tau >0$. The point is that these monodromies generate the monodromy group $\Gamma(4)=\left\{ \left(\begin{array}{cc}
a & b \\ c & d \end{array}  \right)\in {\rm SL}(2,\mathbb{Z}),b=0 \, {\rm mod}\, 4 \right\} $ which is the subgroup of the modular group ${\rm SL}(2,\mathbb{Z})$. The quantum moduli space of the effective theory is now recovered as ${\cal M}=H^+/\Gamma(4)$, where $H^+$ is the upper complex half-plane. Such ${\cal M}$ is precisely moduli space of some toroidal Riemann surface (elliptic curve). This is SW curve and is given by: $y^2(x,u)=(x^2-u^2)-\Lambda^4$ (the family of curves parameterized by the $u$-plane). Now $\tau (u)$ is the following ratio of periods $\frac{\overline{\omega}_D(u)}{\overline{\omega}(u)}$ where $\overline{\omega}_D(u)=\int_{B}\omega$, $\overline{\omega}_D(u)=\int_{A}\omega$ and $\omega=\frac{1}{2\pi}\frac{{\rm d}x}{y(x,u)}$ and $A,B$ generates the homology of the SW curve.
Because of the relation $\tau=\frac{\partial a_D}{\partial a}$ we have $\overline{\omega}_D(u)=\frac{\partial a_D(u)}{\partial u}$ and $\overline{\omega}(u)=\frac{\partial a(u)}{\partial u}$ which means that one recovers $a(u),a_D(u)$ by the integrations on $u$ of the periods. Similarly, the prepotential is given by the following integral ${\cal F}=\int {\rm d}a\, a_D(a)$. 

Introducing the meromorphic 1-form, SW-differential $\lambda_{SW}=\frac{1}{\sqrt{2}\pi}\frac{x^2\, {\rm d}x}{y(x,u)}$, the expressions above read $a_D(u)=\int_B=\lambda_{SW},\, a(u)=\int_A\lambda_{SW}$.

One can also consider the SW curve as $y^2=(x^2-\Lambda^2)(x-u)$ in which case the monodromies read 
\begin{equation}\label{A-4} \begin{array}{cc} M_{\infty}= \left(\begin{array}{cc}
-1 & 2 \\ 0 & -1 \end{array}  \right)\, , \; \; M_{+\Lambda^2}= \left(\begin{array}{cc}
1 & 0 \\ -2 & 1 \end{array} \right)\, ,  \\[5pt] M_{-\Lambda^2}= \left(\begin{array}{cc}
-1 & 2 \\ -2 & 3 \end{array} \right) . \end{array} \end{equation}
They generate the $\Gamma(2)$ subgroup of ${\rm SL}(2,\mathbb{Z})$ which is used when SW theory is embedded in string theory compactified on local CY (see Sec. \ref{strings-1}). The integration of $\lambda$ on $u$-plane gives now: $a(u)=\frac{\sqrt{2}}{\pi}\int_{-\Lambda}^{\Lambda}\frac{{\rm d}x\sqrt{x-u}}{\sqrt{x^2-\Lambda}}$ and $a_D(u)=\frac{\sqrt{2}}{\pi}\int_{\Lambda}^{u}\frac{{\rm d}x\sqrt{x-u}}{\sqrt{x^2-\Lambda}}$. Now, one can consider the moduli space ${\cal M}$ as the $u$-plane with singular points at $\pm \Lambda,+\infty$ (and with $\mathbb{Z}_2$ symmetry). The coordinates $(a,a_D)^T$ is the section of the flat ${\rm SL}(2,\mathbb{Z})$-bundle over ${\cal M}$. This bundle has monodromies as above around the singular points.

Due to the $U(1)_{{\cal R}}$ chiral anomaly, present in the microscopic and effective ${\cal N}=2$ theories, one determines the 1-loop perturbative prepotential from $\frac{\partial^2 {\cal F}}{\partial^2 a}(e^{2i\alpha a})=\frac{\partial^2 {\cal F}}{\partial^2 a}(a)-\frac{4\alpha}{\pi}$ as: ${\cal F}_{1-loop}=\frac{i}{2\pi}a^2 {\rm ln}\frac{a^2}{\Lambda^2}$. There are no other perturbative higher order corrections. Still, non-perturbative, instanton corrections to prepotential are in order. Together, the complete prepotential reads ($k$ is the instanton number): 
\begin{equation}\label{A-5}
{\cal F}= \frac{i}{2\pi}a^2 {\rm ln}\frac{a^2}{\Lambda^2}+\sum_{k=1}^{\infty}{F}_k \left(\frac{\Lambda}{a}\right)^{4k}  . 
\end{equation}


\begin{thebibliography}{10}


\bibitem{AssKrol2010ICM}
T.~Asselmeyer-Maluga, J.~Kr{\'o}l, Small exotic smooth $R^4$ and string theory,
 in {\em International Congress of Mathematicians ICM, Hyderabad, India 2010, Short Communications Abstracts Book}, R. Bathia (Ed.), Hindustan Book Agency, p. 400 (2010).

\bibitem{AsselmKrol2011d}
T.~Asselmeyer-Maluga, J.~Kr{\'o}l, {\em Int. J. Geom. Meth. Mod. Phys.}, \textbf{9}, 2012, arXiv:1102.3274.

\bibitem{AsselmeyerKrol2011}
T.~Asselmeyer-Maluga, J.~Kr{\'o}l, {\em Int. J. Mod. Phys. A} \textbf{26}, 1375, 2011, arXiv:1101.3169.

\bibitem{AsselmeyerKrol2011b}
T.~Asselmeyer-Maluga, J.~Kr{\'o}l, {\em Int. J. Mod. Phys. A} \textbf{26}, 3421, 2011, arXiv:1105.1557.

\bibitem{QG-2012}
J.~Kr\'ol, Quantum gravity insight from smooth 4-geometries on trivial $\mathbb{R}^4$, in \emph{Quantum Gravity}, Rodrigo
  Sobreiro (Ed.), ISBN: 978-953-51-0089-8, InTech, Available from:
  http://www.intechopen.com/articles/show/title/quantum-gravity-insights-from-smooth-4-geometries-on-trivial-r4, 2012.

\bibitem{Krol2011d}
J.~Kr{\'o}l, {\em Acta. Phys. Pol. B} \textbf{42}(11), 2343, 2011.
  

\bibitem{Krol2010}
J.~Kr{\'o}l, {\em Ann. Phys. (Berlin)} \textbf{19}(3), 2010.

\bibitem{BCOV1993}
M.~Bershadsky, S.~Cecotti, H.~Ooguri and C.~Vafa, {\em Commun. Math. Phys.} \textbf{165}, 311, 1994, arXiv:hep-th/9309140v1.

\bibitem{Klemm1997}
S.~Katz, A.~Klemm and C.~Vafa, {\em Nucl. Phys. B} \textbf{497}, 173, 1997, \newblock arXiv:hep-th/9609239v2.

\bibitem{Nekr2002}
N.~A. Nekrasov, {\em Adv. Theor. Math. Phys.} \textbf{7}, 831, 2004, arXiv:hep-th/0206161.

\bibitem{Klemm2009}
M.~Huang and A.~Klemm, Holomorphicity and modularity in Seiberg-Witten theories with matter, {B}onn-TH-09-01,CERN-PH-TH/2008-236, 2009, arXiv:0902.1325v1.

\bibitem{Klemm2007b}
M.~Huang and A.~Klemm, {\em JHEP} \textbf{0709}, 054, 2007, arXiv:hep-th/0605195v2.

\bibitem{Klemm2006}
M.~Aganagic,
  V.~Bouchard and A.~Klemm, {\em Commun. Math. Phys.} \textbf{277}, 771, 2008, arXiv:hep-th/0607100v2.

\bibitem{Witten1993}
E.~Witten, {Q}uantum background independence in string theory, arXiv:hep-th/9306122.

\bibitem{Lerche1996}
W.~Lerche, {\em Fortsch.Phys.} \textbf{45}, 293, 1997, arXiv:hep-th/9611190.

\bibitem{Nekr2003}
N.~A. Nekrasov and A.~Okounkov, {S}eiberg-{W}itten theory and random partitions, arXiv:hep-th/0306238.

\bibitem{Min2012}
M.~Huang, {O}n Gauge Theory and Topological String in Nekrasov-Shatashvili Limit, IPMU 12-0093, 2012, arXiv:1205.3652.

\bibitem{Sierakowski}
A.~Sierakowski, Ph.D. thesis, University of Copenhagen - Department of Mathematical
  Sciences, 2009.

\bibitem{ConnesMosc2004}
A.~Connes, and H.~Moscovici, {\em Moscow Math. J.} \textbf{4}, 2004.

\bibitem{AsselmKrol2012b}
T.~Asselmeyer-Maluga and J.~Kr\'ol, {T}owards superconformal and
  quasi-modular representation of exotic smooth $\mathbb{R}^4$ from superstring
  theory I, 2012.

\bibitem{Phillips2007}
N.~C. Phillips, {O}ttawa {S}ummer {S}chool course on
  crossed product $C^{\star}$ algebras, 2007.

\bibitem{Zagier2008}
D.~Zagier, Elliptic modular forms and their applications, in {\em The 1-2-3 of Modular Forms}, Springer, 2008. 

\bibitem{Sladkowski2001}
J.~S{\l}adkowski, {\em Int.J. Mod. Phys. D} \textbf{10}, 311, 2001.

\bibitem{Asselm-Krol-2011}
T.~Asselmeyer-Maluga, and J.~Kr\'ol, {E}xotic Smoothness and Quantum Gravity II: exotic R4, singularities and cosmology, arXive:1112.4882.

\bibitem{Witten1994}
E.~Witten, {\em J.Math.Phys.} \textbf{35}, 5101, 1994, arXiv:hep-th/9403195.

\bibitem{Taubes1997}
M.~Hutchings and C.~H. Taubes, in {\em Symplectic geometry and topology},  Y.~Eliashberg, L.~Traynor, (Eds.), Amer. Math. Soc., 1999.

\bibitem{Seiberg1988}
N.~Seiberg, {\em Phys. Lett.} \textbf{206B}, 75, 1988.

\bibitem{Witten-Seiberg1994}
N.~Seiberg and E.~Witten, {\em Nucl.Phys.} \textbf{B426}, 19, 1994, arXiv:hep-th/9407087.

\bibitem{Witten1995}
E.~Witten, {\em Selecta Math.} \textbf{1}, 383, 1995, arXiv:hep-th/9505186.

\bibitem{AsselmKrol2011f}
T.~Asselmeyer-Maluga, P.~Gusin and J.~Kr\'ol, {T}he modification of the energy spectrum of charged particles by exotic open 4-smoothness via superstring theory, will appear in \emph{Int. J. Geom. Meth. Mod. Phys.} \textbf{10}(1), 2013, arXiv: 1109.1973.

\bibitem{Bilal1996}
A.~Bilal, {D}uality in N=2 SUSY SU(2) Yang-Mills Theory: A pedagogical introduction to the work of Seiberg and Witten, 1996, arXiv:hep-th/9601007.

\end{thebibliography}

\end{document}